\documentclass[preprints,article,accept,moreauthors,pdftex]{Definitions/mdpi}
\usepackage{url}
\firstpage{1} 
\pubvolume{xx}
\issuenum{1}
\articlenumber{5}
\pubyear{2019}
\copyrightyear{2019}
\history{Received: date; Accepted: date; Published: date}

\Title{Development of the first Portuguese radar tracking sensor for Space Debris}

\Author{João Pandeirada $^{1,2}$, Miguel Bergano $^{1,3}$, João Neves $^{4}$, Paulo Marques $^{5}$, Domingos Barbosa $^{1}$, Bruno Coelho $^{1}$ and Valério Ribeiro $^{1,2}$}

\AuthorNames{João Pandeirada, Miguel Bergano and João Neves}

\address{%
$^{1}$ \quad Instituto de Telecomunicações, Aveiro, Portugal; joao.pandeirada@av.it.pt (J.P); jbergano@av.it.pt (M.B); dbarbosa@av.it.pt (D.B); brunodfcoelho@av.it.pt (B.C);  valerio.ribeiro@av.it.pt (V.R)\\
$^{2}$ \quad Universidade de Aveiro, Aveiro, Portugal\\
$^{3}$ \quad ESTGA - Universidade de Aveiro, Aveiro, Portugal\\
$^{4}$ \quad CINAV - Escola Naval, Alfeite, Portugal; fidalgo.neves@marinha.pt (J.N)\\
$^{5}$ \quad Instituto de Telecomunicações / ISEL-IPL, Lisboa, Portugal; pmarques@isel.pt (P.M)
}

\abstract{Currently, space debris represents a threat for satellites and space-based operations, both in-orbit and during the launching process. The yearly increase in space debris represents a serious concern to major space agencies leading to the development of dedicated space programs to deal with this issue. Ground-based radars can detect Earth orbiting debris down to a few square centimeters and therefore constitute a major building block of a space debris monitoring system. New radar sensors are required in Europe to enhance capabilities and availability of its small radar network capable of tracking and surveying space objects and to respond to the debris increase expected from the New Space economy activities. This article presents ATLAS, a new tracking radar system for debris detection located in Portugal. It starts by an extensive technical description of all the system components followed by a study that estimates its future performance. A section dedicated to waveform design is also presented, since the system allows the usage of several types of pulse modulation schemes such as LFM and phase coded modulations while enabling the development and testing of more advanced ones.
By presenting an architecture that is highly modular with fully digital signal processing, ATLAS establishes a platform for fast and easy development, research and innovation. The system follows the use of Commercial-Off-The-Shelf technologies and Open Systems which is unique among current radar systems.}


\begin{document}

\section{Introduction}
The use of sensors (which include telescopes, lasers and radars) has been a growing area of interest for monitoring the space environment around Earth. Due to the growing number of space debris, there is a need to predict and adjust the orbits of the more than 2000 active satellites \cite{statista}, avoiding collisions with other inactive satellites or debris, to guarantee their long-term operation and investment.

These actions are core activities of the Space Surveillance and Tracking (SST) concept or, in a broader sense, are framed within the Space Situational Awareness (SSA) domain, notwithstanding the different concepts acknowledged between a USA concept, strongly related to the Defense, or the European concept, strongly related to a dual-use or purely civil use compliant with the European Union (EU) or the European Space Agency (ESA) guidelines.

The Earth near-space environment has been getting polluted with space debris since the onset of the space exploration era in 1957. The envisaged space launching actions to put more satellites in orbit including the announced mega constellations or broadband internet satellites, e.g. 
Starlink Space X and One Web, is a growing concern which drives the need to create, maintain and improve accurate sensor networks to collectively contribute to a safe use of space. The safety and security of space assets is an important matter of national interests and central to international treaties and regulations, namely those agreed to in the framework of the United Nations, at the Committee on the Peaceful Uses of Outer Space \cite{outerspace}.

Sensors networks comprise both ground based and space based assets, although only a few nations can afford to have space based ones. Regarding the ground based sensors, these are usually divided in three main groups: telescopes, radars and lasers. 

Radar usage for SST and SSA purposes is increasing due to its capacity for monitoring the space environment during day and night, and insensitivity both to light pollution and to weather conditions. A tracking radar distinguishes itself not only by recognizing the presence of a target and determining a target location in range and in angle coordinates, but also because it is able to follow the target and observe it over some time range thus improving the information accuracy about its trajectory \cite{skolnik2001}.

The principle of tracking radar is to use the angular error signal, defined by the difference between the target direction and the reference direction given by the axis of the antenna, to adjust the antenna’s pointing direction. The tracking radar attempts to position the antenna with zero angular error (i.e., to locate the target along the reference direction) \cite{Chen2004}	.

The first developments of tracking radars date back from World War II \cite{Galati2016}  but their biggest developments date from the Cold-War \cite{Lemnios2000}. During those times, both the USA and the Soviet Union built large networks of ground-based tracking facilities that collected data on man-made objects in Earth orbit. They consisted, primarily, of large tracking radars and optical telescopes, often dual-purposed with other military missions, such as ballistic missile warning and tracking. The main focus of space surveillance at that time was on determining the precise orbit of man-made objects around the Earth \cite{Gianopapa2015}. Later, they were improved to make them useful for Combat Weapon Systems, such as the Thales STIR-Tracking and Illumination radar \cite{Thales2020}.

The tracking of distant objects in the higher layers of the atmosphere, in the lower orbits around Earth or even far distant objects throughout the galaxy, continued to develop up to complex phased array systems such as the Square Kilometer Array (SKA) radio telescope \cite{SKA2020} or the  brand new US Space Fence, costing several million dollars \cite{EOPORTAL2020}, to name a few. Tracking radars are now suitable for Low-Earth Orbit (LEO) tracking or space object catalogue accuracy improvement due to electronic components developments and increased digital processing capability.

Referring to existing operational tracking radars for in-orbit object detection (and disregarding the ones based on phased arrays, which usually also have surveillance capability), these can be found all over the world and with different capabilities.

On the US side, the development of these radars has a long history with the development of several radars, most of them for the Ballistic Missile Defense program, supported by DARPA (Defense Advanced Research Projects Agency), which were closely related to the detection of objects in orbit \cite{Lemnios2000}. Notwithstanding the huge variety of radars that were developed and used (some of them have been decommissioned), two of the most interesting and relevant radars used for Space Debris detection, tracking and characterization are the Haystack and the HAX (Haystack Auxiliary Radar) radars. The HAX radar has a 12.2m antenna, operating at 16.7GHz and is much smaller than Haystack which has a 36.6m antenna operating at 10GHz \cite{Stokely2006} , both in a Cassegrain configuration. HAX had twice the bandwidth of Haystack, being capable of producing correspondingly sharper images of near-earth orbit objects. The cost of building HAX, which became operational in 1994, was reduced by using a refurbished antenna and by sharing signal processing and data processing systems with the Haystack Long Range Imaging Radar (LRIR) (means that the two radars could not operate simultaneously).  The Haystack radar has been operational since 1964, but from 2010 to 2014 it underwent several upgrades becoming the Haystack Ultra-Wideband Satellite Imaging Radar (HUSIR), able to also operate in the W-band with a bandwidth up to 8GHz, which allowed the HAX to detect objects down to 3cm and the HUSIR down to 3mm at an altitude up to 1000km \cite{Kennedy2019}.

The Space Debris networks around the world includes facilities from other major players, such as China \cite{china} and Russia and other spacefaring nations such as India \cite{india} \cite{india1}, Iran, Brazil \cite{latinamerica} \cite{chinabrazil}, Japan or North Korea. However, most of those nations projects’ and infrastructures are linked to the military and involved in deep secrecy. It’s highly probable that radio-telescopes and other related ground-based sensors support missions including intelligence collection, counterspace targeting, ballistic missile early warning, spaceflight safety, satellite anomaly resolution, and also space debris tracking \cite{DIA2019}.

In the European continent the development of tracking radars for orbit objects detection has evolved for some time and nowadays there are around 50 different radio telescopes. Just to mention some which have played a relevant role in space debris and satellite detection, it is worth mentioning the German Tracking and Imaging Radar (TIRA) system, with a 34 meter antenna, located near Bonn. In a monostatic configuration it is able to detect objects of diameters down to 2 cm at 1000 km, but in a bistatic configuration, in synergy with the Effelsberg Radio Telescope of the Max Planck Institute for Radio Astronomy, which has a 100 meter dish antenna, the system can detect objects down to 1 cm diameter \cite{Mehrholz1997}  \cite{Jehn1999}. The TIRA radar has been one of the most useful radars in Europe in the SST context and its accuracy and imaging capabilities, along its continued upgrades, made it a valuable asset. 
The Sardinia Radio Telescope (SRT) is an Italian radio telescope facility with recent radar capabilities with a 64 meter antenna located in the Sardinia island, operating only since 2012. Compared to equivalently sized radars, it has a fairly good speed in azimuth and elevation tracking (0.85º/s and 0.5º/s, respectively) and is able to operate from 300MHz up to 15 GHz. It is intended to work in a bistatic configuration, using the Flight Terminator System (FTS), about 40 km apart from the first, which configuration is named BIRALET (BIstatic RAdar for LEo Tracking) \cite{Muntoni2017}. BIRALET has been used in the context of the European SST for LEO detection, but its capabilities, namely the SRT antenna, envisage the possibility to go beyond this orbit.

On the French side, one of the countries in Europe that has more investments in the space segment, there have been several developments in radars, but the most relevant and interesting of them are in the military side. The French space surveillance network comprises a set of three tracking radars named SATAM, in three different locations (one of it with deployable capability). This configuration is not primarily dedicated to space surveillance, but is used for space events, detection of risk collision and atmospheric reentries, being its speed (about 40º/s) an incomparable added-value, notwithstanding the lower dimension of the dishes, compared with the previous ones. 

The European Space Surveillance and Tracking (EUSST) is a support framework setup by the European Commission to create an autonomous monitoring network supporting the detection and tracking of space debris and issue alerts when evasive actions may be necessary. EUSST aims to establish an SST capability at the European level with an appropriate level of European autonomy \cite{EUNION2014},\cite{eusst}. Portugal is a member of the EUSST program and is developing capabilities both in optical and radar~sensors. 

This paper describes the development of the first Portuguese tracking sensor for space debris and is organized as follows. Section 1 explains the necessity of monitoring the space environment and how nations worldwide have tackled this problem throughout the years. It also provides examples of already deployed tracking radars with similar functions to the current one in development. Section 2 introduces ATLAS (r\textbf{A}dio \textbf{T}e\textbf{L}escope p\textbf{A}mpilhosa \textbf{S}erra) and provides a detailed technical description of its components: antenna, transmitter, receiver, clock generation, data acquisition and controllers. Section 3 consists of simulation studies that use the technical characteristics of the system in order to estimate some performance metrics: minimum detectable target size, minimum elevation, number of expected observable objects and maximum simultaneous trackable targets. Section 4 explains the importance of waveform design on radar systems and provides some examples of waveforms that can be used by ATLAS. Finally, Section 5 concludes the article and reveals the next steps on the realization of this project.

\section{Implemented System}
\begin{figure}[h]
	\begin{center}
		\includegraphics[width=\columnwidth]{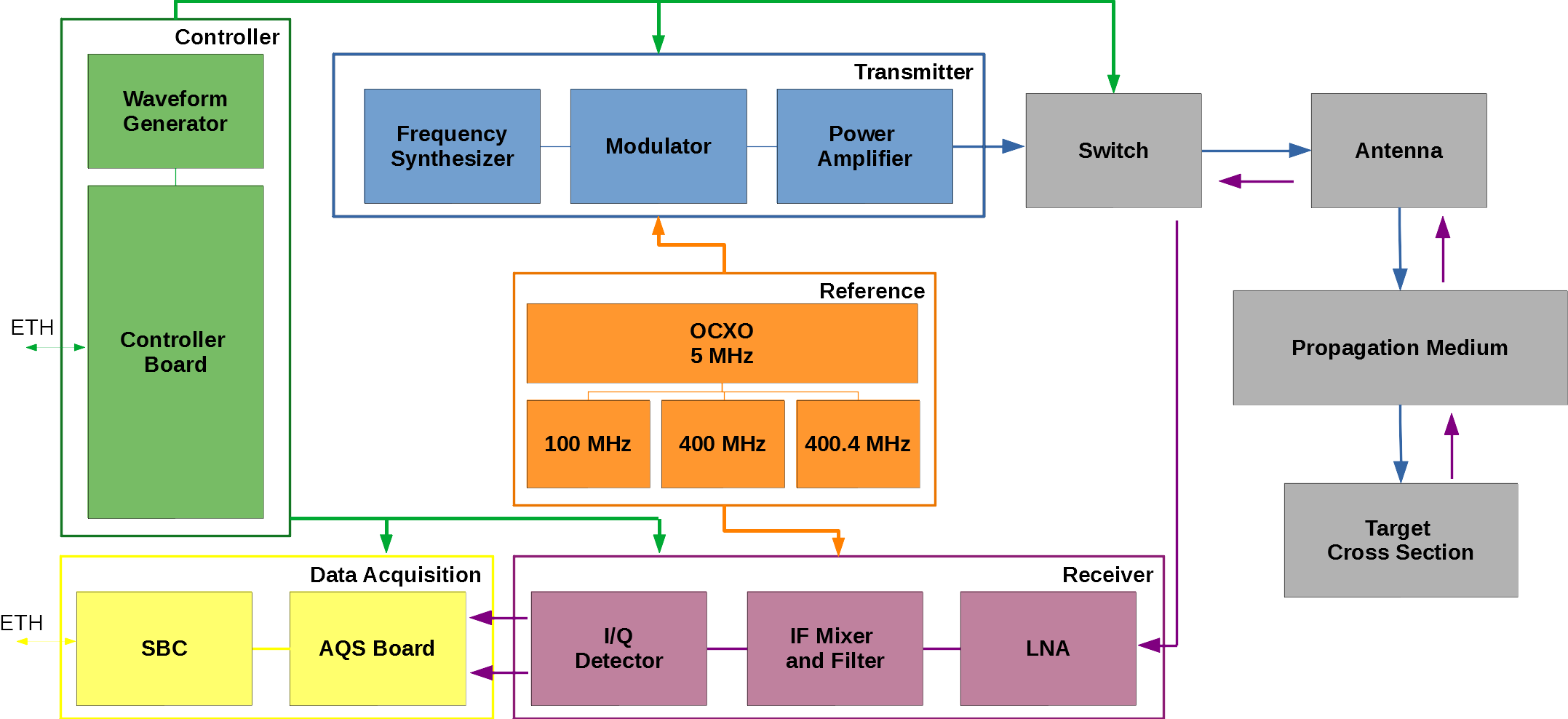}
		\caption{ATLAS system block diagram}
		\label{sysblocks}
	\end{center}
\end{figure}

In order to support and consolidate the pooling of national resources to prepare for the establishment of a SST sensor network, Instituto de Telecomunicações initiated in 2019 a major upgrade of its Cassegrain Antenna located in Pampilhosa da Serra, Portugal. Attached to the antenna will be a complete ground-based radar system, named ATLAS, operating in the C-band (5.56 GHz).

ATLAS is a monostatic pulse radar using solid state power amplifiers (GaN) with a peak output power of 5 kW for tracking debris objects in LEO. Regarding the receiver side, the system is fully coherent with detection and processing fully in the digital domain with a bandwidth of 50 MHz and with the capacity to detect Doppler velocities up to 10.79 km/s. The architecture, depicted in Figure \ref{sysblocks}, is highly inspired in the HAX/Haystack system \cite{husir} developed by NASA , which is a very successful ground-based radar used for debris detection and tracking. The system follows the use of Commercial-Off-The-Shelf (COTS) technologies and Open Systems (OS) which enables a major downstream cost reduction in the development and maintenance of the system \cite{rosa}. Table \ref{soatable} provides a comparison between ATLAS and other tracking radars currently used in SSA activities.

\begin{table}[h]
\caption{Comparison between ATLAS and other tracking radars}
\centering
\label{soatable}
\begin{tabular}{c|c|c|c|c|c|}
\cline{2-6}
                                                   & \textbf{ATLAS} & \textbf{TIRA} & \textbf{BIRALET}       & \textbf{HAX} & \textbf{Haystack} \\ \hline
\multicolumn{1}{|c|}{\textbf{Operating Frequency}} & 5.56 GHz          & 1.333 GHz     & 410 MHz                & 16.7 GHz     & 10 GHz            \\ \hline
\multicolumn{1}{|c|}{\textbf{Peak power}}          & 5 kW              & 1 MW          & 4 kW                   & 50 kW        & 250 kW            \\ \hline
\multicolumn{1}{|c|}{\textbf{Waveform type}}       & Pulsed            & Pulsed        & Continuous             & Pulsed       & Pulsed            \\ \hline
\multicolumn{1}{|c|}{\textbf{Antenna Gain}}        & 46 dB             & 49.7 dB       & 13 dB (TX), 47 dB (RX) & 63.64 dB     & 67.23 dB          \\ \hline
\multicolumn{1}{|c|}{\textbf{Antenna Beamwidth}}   & 0.73º             & 0.50º         & 30º (TX), 0.8º (RX)    & 0.1º         & 0.058º            \\ \hline
\multicolumn{1}{|c|}{\textbf{Receiver Bandwidth}}  & 80 MHz            & 250 kHz       & 5 MHz                  & 1 MHz        & 1 MHz             \\ \hline
\multicolumn{1}{|c|}{\textbf{Topology}}            & Monostatic        & Monostatic    & Bistatic               & Monostatic   & Monostatic        \\ \hline
\end{tabular}
\end{table}

\subsection{Radar site and Antenna}

We obtained a Vertex RSI C band, 9 metre Cassegrain Antenna originally used for Satellite Communications in the Azores Terceira Island, Portugal. The antenna was moved to Pampilhosa da Serra, Portugal, a site with a low Radio Frequency Interference (RFI) environment \cite{rui}. New foundations, electrical power and lightning protection were installed at this new site. The mounting pedestal underwent a major adaptation in order to comply with a fast, continuous azimuth antenna rotation. Originally intended for radio astronomy surveys, this facility represents a step forward in our technological development and evolution of high-performance space radar instrumentation systems. 

The major improvements have given the station space telemetry and tracking capabilities. These improvements will represent one of Portugal’s main support pilot facilities for space monitoring and will provide support to current and future space missions. The upgrade approach manages to automatically, and quickly, attain the best antenna steering strategy to successfully track the objects.

The data output capabilities have been upgraded for wide bandwidth. These capabilities are now available to users on site and remotely via a high throughput connection. Extensive user support will be provided for these new facilities, including assistance, monitoring and control, antenna interfaces and output interfaces. Operation and data reduction is also available for object tracking.

The antenna is depicted in Figure \ref{antenna} and specifications are detailed in Table \ref{antennatable}.

\begin{table}[]
\begin{minipage}[b]{0.5\linewidth}
\captionof{table}{Antenna specifications}
\centering
\label{antennatable}
\begin{tabular}[b]{|c|c|}
\hline
\textbf{Optical Configuration}            & Cassegrain                 \\ \hline
\textbf{Mount configuration}              & Alt-Azimuthal              \\ \hline
\textbf{Primary aperture}                 & 9.0 m                      \\ \hline
\textbf{Primary depth}                    & 1.53                       \\ \hline
\textbf{Secondary aperture}               & 1.17                       \\ \hline
\textbf{F/D}                              & 0.368                      \\ \hline
\textbf{Surface rms (static)}             & 0.5 mm                     \\ \hline
\textbf{Azimuth Travel}                   & 0º to 360º                 \\ \hline
\textbf{Azimuth Travel Rate}              & 4 rpm                     \\ \hline
\textbf{Elevation Travel}                 & 35º to 90º                 \\ \hline
\textbf{Elevation Travel Rate}            & 2º/s                       \\ \hline
\textbf{Maximum Operational Wind (5GHz)}  & 25 km/h                    \\ \hline
\textbf{Survival Winds}                   & 150 km/h                   \\ \hline
\textbf{Reflector Weight}                 & 1900 kg                    \\ \hline
\textbf{Pedestal Weight}                  & 3500 Kg                    \\ \hline
\textbf{Foundation Size}                  & 4.5 x 4.5 x 0.75 m         \\ \hline
\textbf{Concrete Volume}                  & 12 ms                      \\ \hline
\textbf{Beamwidth @ 5 GHz}                & 44 arcmin                 \\ \hline
\textbf{Beamwidth @ 10 GHz}               & 22 arcmin                 \\ \hline
\textbf{Pointing Accuracy (wind limited)} & \textless{}1/20 beamwidth  \\ \hline
\textbf{Gain @ 5 GHz ($G$)}                     & 46 dB                      \\ \hline
\textbf{G/T (5 GHz)}                       & 39 dB/K                    \\ \hline
\end{tabular}
\end{minipage}\hfill
\begin{minipage}[b]{0.41\linewidth}
\centering
\includegraphics[width=\linewidth]{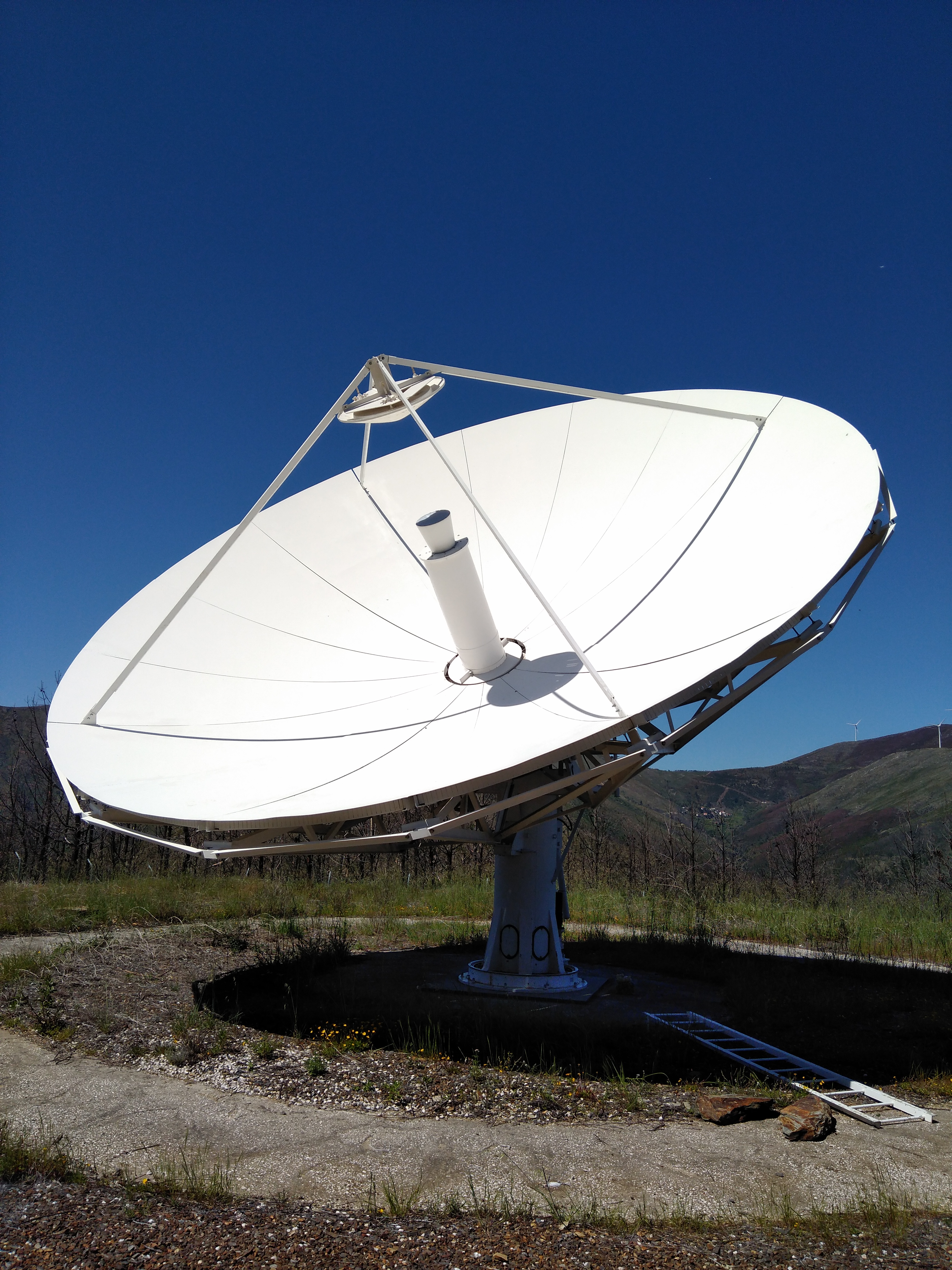}
\captionof{figure}{ATLAS Antenna}
\label{antenna}
\end{minipage}
\end{table}


\subsection{Transmitter}
Active radars require Power Amplifiers (PAs), which should be small, efficient and low-cost. We implemented a C-band PA centered at 5.56GHz with an output peak power of 5 kW (67dBm) from a 30dBm drive with a Power Added Efficiency (PAE) of 67 \%.
GaN technology is suitable for these requirements and can do so with high efficiency. The PA was implemented by readily available COTS technology.
Compact Radio Frequency (RF) power sources now exist with the recent release of high-power solid-state RF amplifiers such as the Wolfspeed/Cree CGHV59350 high-electron-mobility transistors (HEMTs) \cite{cree}. These HEMTs are capable of 500 W of RF power each. Our design follows the traditional approach of combining a large number of solid-state RF amplifiers into a high-power all-solid-state RF system \cite{simons}.
The transmitter, depicted on Figure \ref{sysblocks} in blue, is responsible for outputting the RF signal. All the relevant operating parameters are summarized in Table \ref{txtable}. The signal generation starts with a frequency synthesizer at 5.160 GHz which is fed to a frequency translator. Along with the 5.160 GHz signal, the frequency translator receives an intermediate frequency of 400 MHz from the reference module and generates the desired 5.560 GHz signal. This signal is fed into a PIN modulator (PIN diode based) responsible for modulating the 5.560 GHz carrier in amplitude. The PIN modulator is also connected to the controller board in order to receive the user-defined pulse shape. The modulated signal goes to a driver, to achieve a 30dBm minimum signal to be provided to the PA. We selected the previously mentioned Cree HEMTs at 5–6 GHz because these HEMTs provide the highest available RF power as required (67dBm).
The target performance requirements for the amplifier are listed in Table \ref{txtable}.

\begin{table}[h]
\caption{Transmitter Operating Parameters}
\centering
\label{txtable}
\begin{tabular}{|c|c|}
\hline
\textbf{Peak power}            & 5 kW                                     \\ \hline
\textbf{Transmitter Frequency ($1/\lambda$)} & 5.56 GHz                                 \\ \hline
\textbf{Waveform}              & Arbitrary amplitude modulation           \\ \hline
\textbf{Max. Pulse length ($\tau$)}     & 10 s                                     \\ \hline
\textbf{Phase Noise}           & -91.3dBc{[}Hz{]} @ 100 kHz               \\ \hline
\textbf{Modulator}             & Modified D195 \cite{pinmod}                  \\ \hline
\textbf{PA Transistors}        & CGHV59070 \cite{gan1}, CGHV59350 \cite{cree} \\ \hline
\end{tabular}
\end{table}

\subsection{Receiver}
The antenna is equipped with a corrugated feed horn, followed by a  polarizer  and  by  an  orthomode transducer (OMT) to carry the left and right circular polarizations (LHCP for Tx and RHCP for Rx). A waveguide to coaxial transition performs the connection to the receiver. The receiver has a limiter and switch right at the input that handles an excess peak power of 20 W for about 10 ms and blocks the signal during transmission with an attenuation of 30 dB.

The receiver, depicted on Figure \ref{sysblocks} in purple, includes an analog chain that is responsible for amplifying the signals (target echoes) to the required digital entry levels of the acquisition board (yellow). The receiver starts with an low noise amplifier (LNA), internally developed, using GaAs technology, presenting a noise factor of 0,7 dB at 5.56GHz. Next, there is a local oscillator at 5.160 GHz to convert it down to a 400 MHz intermediate frequency (IF). The local oscillator is provided by the same reference signal from the transmitter, guaranteeing a coherent conversion. The IF then passes through an IF filter and a digital variable gain amplifier (DVGA) which is connected to the controller~board.

Finally, the I/Q detector converts the signal to base-band (or 400 kHz) and outputs the I and Q components of the echo signal. The conversion to 400 kHz is used for accommodating Doppler shifts, providing velocity measurements up to 10.79 km/s. Table \ref{rxtable} summarizes all relevant parameters of the receiver chain.

\begin{table}[h]
\caption{Receiver Operating Parameters}
\centering
\label{rxtable}
\begin{tabular}{|c|c|}
\hline
\textbf{IF}                                 & 400 MHz       \\ \hline
\textbf{LNA Noise figure (\textless{}15ºC) ($N_f$)} & 0.7 dB        \\ \hline
\textbf{Receiver Temperature ($T$)}               & 15ºC          \\ \hline
\textbf{IF Filter BW ($B$)}                       & 80 MHz @ -3dB \\ \hline
\textbf{IQ Detector output BW}              & 50 MHz       \\ \hline
\end{tabular}
\end{table}

\subsection{Reference}
The reference module, depicted on Figure \ref{sysblocks} in orange, generates reference signals at 100, 400 and 400.4 MHz. The main reference oscillator is a 5 MHz oven-controlled crystal oscillator (OCXO) from MTI \cite{ocxo}. The 5 MHz reference locks a pair of 100 MHz XPLO's that are then multiplied by 4 to generate the 400 and 400.4 MHz. This module is connected to the controller board in order to switch between IQ detection at base band or with 400 KHz offset.

\subsection{Controller}
The controller, depicted on Figure \ref{sysblocks} in green, is responsible for the digital control of the whole system. It is composed of a controller board, a pulse generator and an arbitrary waveform generator~(AWG). 

The board establishes a telnet connection with the host PC via Ethernet (ETH) where the user can program the radar using a command line interface (CLI). The CLI can be used to define a great range of radar parameters such as the pulse length, shape, time between pulses, number of pulses and a lot of other internal parameters such as gains and delays. It is also possible to display several system status parameters such as pulse configurations and temperatures. Table \ref{controllertable} shows all the variables that can be monitored and configured.

The pulse generator is responsible for timing all the components of the system. It enables/disables the transmitting and receiving chains and triggers the arbitrary wave generator as well as the acquisition board. This enables the system to be fully coherent and synchronized from signal generation to acquisition.

Finally, the AWG enables the design of arbitrary amplitude modulated waveforms in the digital domain that can be uploaded to the system.

\begin{table}[h]
\caption{Controller board parameters}
\centering
\label{controllertable}
\begin{tabular}{|c|c|}
\hline
\textbf{Waveform resolution}              & 10 ns\\ \hline
\textbf{Pulse repetition frequency ($f_p$)}       & 10 MHz (max)\\ \hline
\textbf{Duty Cycle}                       & 10\% (max)\\ \hline
\textbf{Number of pulses for integration} & Variable\\ \hline
\textbf{Temperature Monitoring}           & \multicolumn{1}{l|}{Tx, Rx, LNC, Driver, PA, Coupler} \\ \hline
\textbf{System Monitoring}                & \multicolumn{1}{l|}{\begin{tabular}[c]{@{}l@{}}Pulse shape file name, Tx/Rx gain,\\ No. Pulses, PRF, Pulse length ...\end{tabular}} \\ \hline
\end{tabular}
\end{table}

\subsection{Data acquisition}
The data acquisition, depicted on Figure \ref{sysblocks} in yellow, contains the acquisition board (AQS) and a single board computer (SBC). The acquisition board is triggered by the controller board and is responsible for converting the I/Q signal components to the digital domain, by employing two 16 bit ADC's at a maximum sample rate of 100 MS/s. The computer (MIO 2360 N-32A1E by Advantech \cite{sbc}) receives the I/Q signal in the digital domain and performs the necessary digital signal processing. Communication between the AQS and the SBC is done via full speed USB 3.0 (100~MS/s) and a Fast/Slow mode buffer is selectable, where a compromise between sample resolution and sample rate can be chosen.

\section{System expected capabilities}
Given the technical specifications of the current system, it is possible to make estimates on how it will perform. These estimates are not only useful for predicting the systems applicability but also to identify its limitations and guide to future upgrades.

\subsection{Minimum detectable target size}
The radar range equation represents the physical dependencies of the transmitted power and is used to obtain the power in the receiving antenna \cite{skolnik}. Using the radar equation, we can obtain the dependence of the signal to noise ratio (SNR) with the radar specifications:

\begin{equation}
    \text{SNR} = \frac{P_{av} G^2 \lambda^2 \sigma n e(n) F^4}{4\pi^3 \tau f_p R^4 N_F k T B L_s }
    \label{eq:radareq}
\end{equation}

where:
\begin{itemize}
    \item $P_{av}$ is the average transmitted power ($W$)
    \item $G$ is the antenna gain
    \item $\lambda$ is the operating wavelength ($m$)
    \item $\sigma$ is the target radar cross section (RCS) ($m^2$)
    \item $n$ is the number of integrated pulses
    \item $e(n)$ is the integration efficiency
    \item $F$ accounts for all the propagation effects
    \item $\tau$ is the pulse width ($s$)
    \item $f_p$ is the pulse repetition frequency ($Hz$)
    \item $R$ is the distance to the object ($m$)
    \item $N_F$ is the noise factor of the receiver
    \item $k$ is the Boltzmann constant ($\frac{J}{K}$)
    \item $T$ is the receiver temperature ($K$)
    \item $B$ is the bandwidth of the receiver ($Hz$)
    \item $L_s$ accounts for all system losses
\end{itemize}

Some of the radar specifications are fixed and have already been shown during this article. Regarding propagation losses it is known that for frequencies below 10 GHz losses due to atmospheric absorption may be neglected \cite{itur}, for that reason we neglected for now the $F$ factor. In future work we will include a detailed modelling of the atmospheric propagation effects including tropospheric absorption. Accounting for all the losses on the waveguide transition from the antenna to the polarizer/orthomode transducer (OMT)/switch and the coaxial transition from the transmitter/receiver to the switch, we obtained $L_s = 0.6$ dB.

By changing the pulse length, pulse repetition frequency and the number of integrated pulses, it is possible to establish a desired SNR for an established target RCS. By doing a statistical analysis on the mean elevation speed of the debris obtained from the Space-Track catalogue (explained in 3.3) and considering our antenna beamwidth, we can safely define 20 pulses for integration. Since the radar is fully coherent with integration in the digital domain, we can assume a coherent integration with full efficiency ($e(n) = 1$). In order to obtain the desired SNR, we defined $\tau = 3.3$ ms and $f_p = 30$ Hz, obtaining the following performance results:
\begin{itemize}
    \item Maximum Unambiguous Range: 5000 km
    \item Maximum Unambiguous Velocity: 0.4 m/s
    \item SNR for a 1 m\textsuperscript{2} RCS at 10\textsuperscript{3} km: 39.55 dB
    \item SNR for a 10 cm\textsuperscript{2} RCS at 10\textsuperscript{3} km: 9.55 dB
\end{itemize}

We can conclude that we can measure targets up to 1000 km in distance with a minimum RCS of~10 cm\textsuperscript{2}.

Since debris objects in LEO have velocities of around 7.8 km/s, obtaining a maximum unambiguous velocity of 0.4 m/s results in inability to measure the Doppler shift. This is due to the fact that we need a low pulse repetition frequency (PRF) for maintaining range unambiguity. It is still possible to measure velocities without Doppler shift by taking several range measurements and inferring velocity from the measurements, which in fact is what is done in an SST system, by processing several individual tracks of the same object.

In order to take advantage of the long idle times between the transmitted and received pulses in a low PRF system, we can use a process called pulse interleaving. Pulse Interleaving is a technique that inserts carefully crafted pulses in the idle time of other pulses. One example is the emission of different waveforms with low cross correlation between them, such as Orthogonal Frequency Division Multiplexing (OFDM) signals, in order to suppress range ambiguity \cite{ofdm_interleaving}.

\subsection{Elevation for initial and final target acquisition}
The location of the radar and its surroundings need to be taken into account for object tracking. ATLAS is located at (40.18ºN,7.87ºW) and is surrounded by mountains, which makes the horizon altitude profile vary in some directions. In these directions, this decreases the arc length of a debris passage thereby reducing its potential tracking time. We thus defined that the minimum elevation for initial and final target acquisition as 30º.

\subsection{Number of expected observable debris objects}
ATLAS was designed to track space objects in LEO orbits below 1000 km of altitude, however, several objects in this range might not be observable due to several constraints:
\begin{itemize}
    \item Objects with an RCS below the minimum detectable threshold;
    \item Orbits with an elevation range below the minimum elevation required for detection and data acquisition;
    \item Orbital speeds exceeding the maximum antenna tracking speed;
\end{itemize}

In order to make an estimate on the number of objects trackable by ATLAS, we retrieved the latest two-line element (TLE) files (which contain the orbital elements) from all the objects categorized as debris with an apogee less than 1000km from the Space-Track public catalogue \cite{spacetrack}. With the orbital elements, we predicted the orbit of each object for the next 7 days (counting from the epoch of the TLE file) with an SGP4 based algorithm \cite{pythonpred}. Then, we intercepted the predicted orbits with the visible portion of the sky for ATLAS, which is obtained by the information provided in Section 3.2. Objects that do not intercept the visible sky are discarded.

\begin{figure}[h]
	\begin{center}
		\includegraphics[width=0.60\columnwidth]{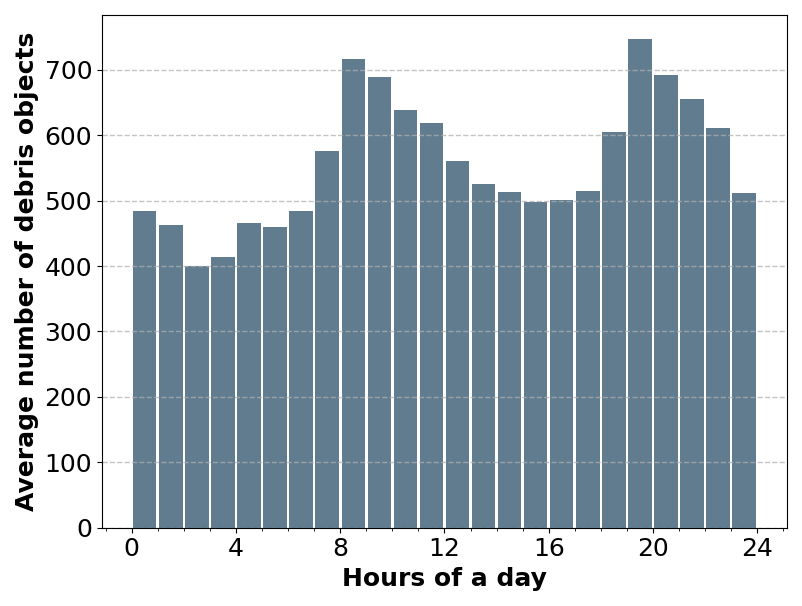}
		\caption{Number of expected observable debris objects during a 24 hour timespan (regardless of RCS).}
		\label{hist1}
	\end{center}
\end{figure}

Since the catalogue does not provide detailed information about the objects RCS, it is not possible to filter out by RCS. All the objects were considered regardless of RCS and the filtering was done by minimum elevation and maximum velocity. Figure \ref{hist1} depicts the number of expected observable objects during a complete day.

\subsection{Maximum simultaneous number of trackable targets}
\begin{figure}[h]
	\begin{center}
		\includegraphics[width=0.60\columnwidth]{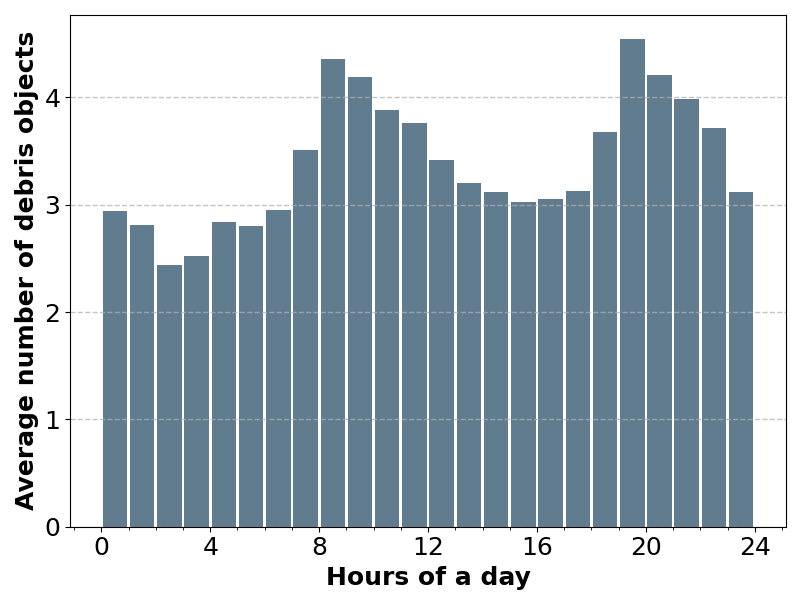}
		\caption{Maximum simultaneous number of trackable targets during a 24 hour timespan.}
		\label{hist2}
	\end{center}
\end{figure}

Since the antenna has a narrow beamwidth (see Table \ref{antennatable}), the amount of sky visible at a specific moment in time is limited and provides an idea on the number of objects illuminated by the radar simultaneously. Figure \ref{hist2} shows the maximum number of simultaneous trackable targets over the course of the day for a beampark configuration. Figure \ref{hist1} and \ref{hist2} share the same shape of the distribution because, if we assume that the number of objects at each bin in Figure \ref{hist1} is uniformly distributed over the visible sky, the number of simultaneous objects is a fraction of that number, thus the shape is retained.

\section{Waveform design}
Designing a proper waveform for the radar pulse is of utmost importance. By manipulating the transmitted pulses in terms of amplitude, frequency and phase it is possible to create waveforms that maximize the range and velocity resolution of the echo signal. In order to evaluate the characteristics of a waveform one can use the ambiguity function (AF). The AF represents the time response of the matched filter when the signal received is affected by a delay $d$ and a Doppler shift $v$ relative to the values expected by the filter and is given by:

\begin{equation}
    |X(d,v)| = \bigg|\int_{-\infty}^{\infty} u(t)u^{*}(t+d)e^{j2\pi v t} dt \bigg|
    \label{eq:acf}
\end{equation}

where $u$ is the complex envelope of the signal \cite{levanon_amb}. By analyzing the zero delay and zero Doppler cuts on the AF it is possible to define the range and velocity resolution. Since ATLAS will work with a low pulse repetition frequency when detecting debris, waveform design focuses on maximizing range resolution instead of velocity.

Matched filtering is the most basic form of signal processing commonly used in radar. It consists in the convolution of the received signal with a filter that is matched with the emitted signal. Matching will result in the maximum attainable SNR at the output of the filter when the signal to which it was matched, after addition of white noise, is passed through it \cite{levanon_matched}.

As defined in Section 3.1, ATLAS will use pulses of 3.3 ms with a pulse repetition frequency of 30 Hz in order to maintain the desired SNR. The waveform used can be arbitrary, within the hardware limits described previously and as long as those power requirements are met.

\subsection{Rectangular Pulse}
\begin{figure}[h]
	\begin{center}
		\includegraphics[width=\columnwidth]{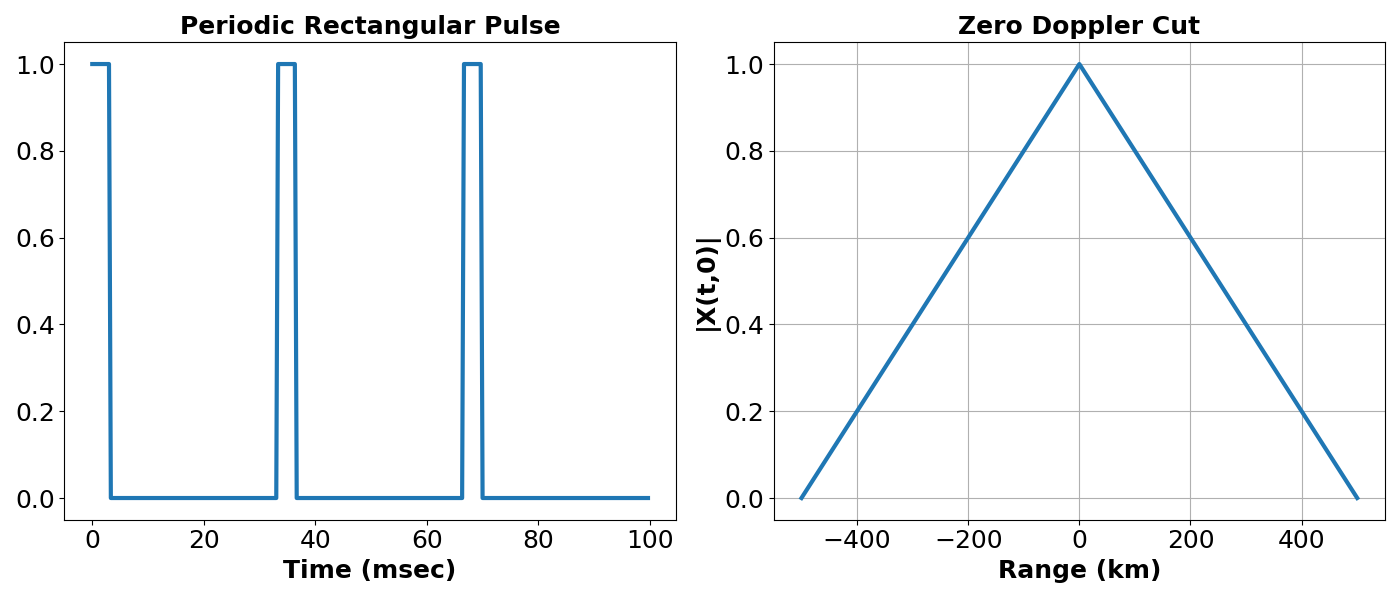}
		\caption{Rectangular pulse (left) and zero cut of AF (right).}
		\label{rect_pulse}
	\end{center}
\end{figure}

Figure \ref{rect_pulse} shows a rectangular pulse of 3.3 ms repeated at 30 Hz and the zero Doppler cut. Using a rectangular pulse results in a range resolution of 500 km which is unacceptable when detecting objects at a maximum of 1000 km.

ATLAS enables the design of arbitrary amplitude modulated waveforms in the digital domain with a resolution of 10 ns. This enables the design of waveforms with high autocorrelation properties that can be used to increase resolution. This signal processing technique is usually called pulse compression \cite{skolnikpulsecomp}.

\subsection{Phase-Coded Pulse (Barker Code)}
The Barker code is a famous binary code used in pulse compression and consists in binary sequences that guarantee that the peak-to-peak sidelobe ratio of the autocorrelation is $M$ where $M$ is the size of the Barker code. Another important feature of the Barker code is that the range resolution increases by $M$ in relation to the rectangular pulse. We can further increase this resolution by using nested Barker codes \cite{levanon_barker}.

\begin{figure}[h]
	\begin{center}
		\includegraphics[width=\columnwidth]{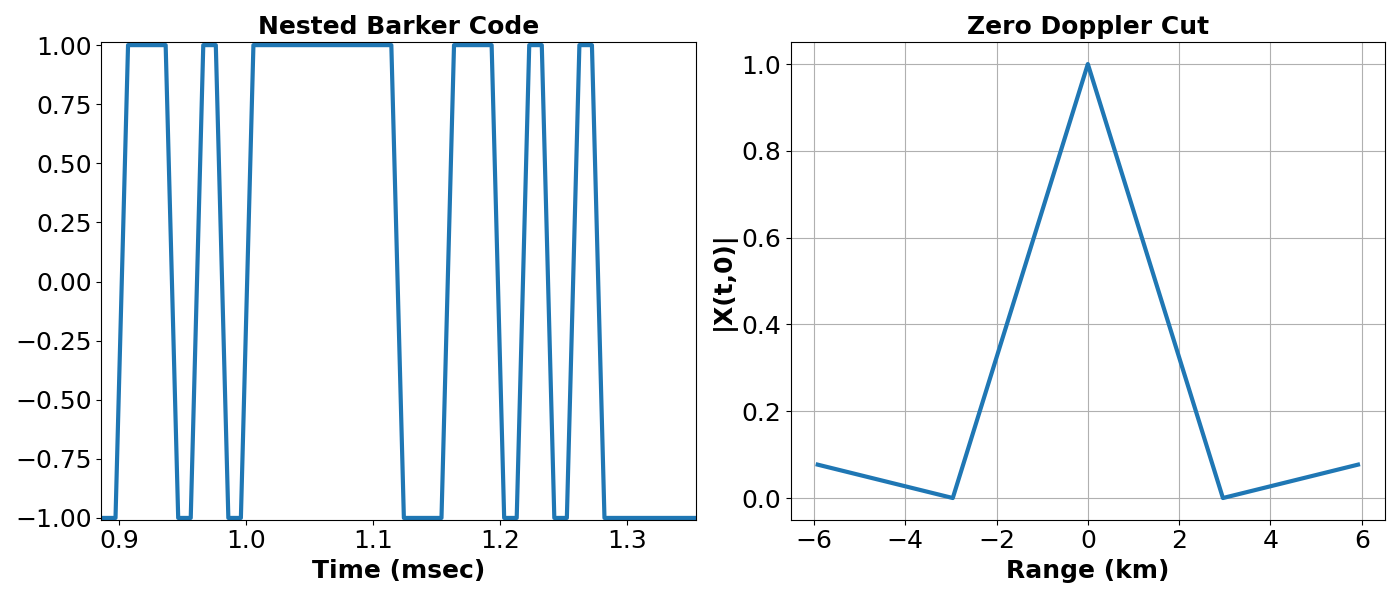}
		\caption{Part of the Nested Barker Code (left) and zero cut of AF (right).}
		\label{nestedbarker}
	\end{center}
\end{figure}

Figure \ref{nestedbarker} shows a Nested Barker code composed of two 13 element codes and the corresponding zero Doppler cut. For visualization purposes, only a small fraction of the code inside of the 3.3 ms pulse is shown. As we can see, the pulse is compressed by a factor of 169, resulting in a range resolution of approximately 2.96 km which is a massive improvement from the rectangular pulse.

Longer binary codes with good auto correlation properties are available which enable even higher compression factors \cite{dzvonkovskaya2008}.

\subsection{Linear Frequency-Modulated Pulse}
Linear frequency modulation (LFM), also known as chirp, is probably the most popular waveform used for pulse compression. It consists in sweeping the carrier wave by the frequency band $B$ during the pulse duration $T$. The complex envelope of a linear chirp is given by \cite{levanon_lfm}:

\begin{equation}
    u(t) = \frac{1}{\sqrt(T)}rect(\frac{t}{T})e^{j\pi k t^2}
    \label{eq:lfm}
\end{equation}

where $k = \pm \frac{B}{T}$ for a linear chirp and the signal indicates an increasing/decreasing sweep. The time-bandwidth product of the signal $B \times T$ is the compression factor in relation to the rectangular pulse, that is, by increasing $B$ the range resolution improves.
Since ATLAS currently only supports amplitude modulation of the carrier, it is not possible to generate traditional chirp pulses because it requires phase modulation. In order to overcome this limitation, we developed an AM Chirp.

The AM Chirp consists in designing a chirp signal at a much lower frequency and using it for modulating the carrier in amplitude. The amplitude modulation is given by:

\begin{equation}
    y(t) = A(t)\cos(2\pi f_c t)
\end{equation}
where $A(t)$ is the modulating signal and $f_c$ is the carrier frequency. In the case of the AM chirp, $A(t)$ is given by:
\begin{equation}
    A(t) = \alpha + \beta \cos[2\pi(ct+f_0)t],~ c = \frac{f_1 - f_0}{T}
\end{equation}

where $\alpha$ and $\beta$ define the modulation parameters, $f_0$ and $f_1$ are the starting and final frequencies and $T$ is the pulse duration. 

\begin{figure}[h]
	\begin{center}
		\includegraphics[width=0.95\columnwidth]{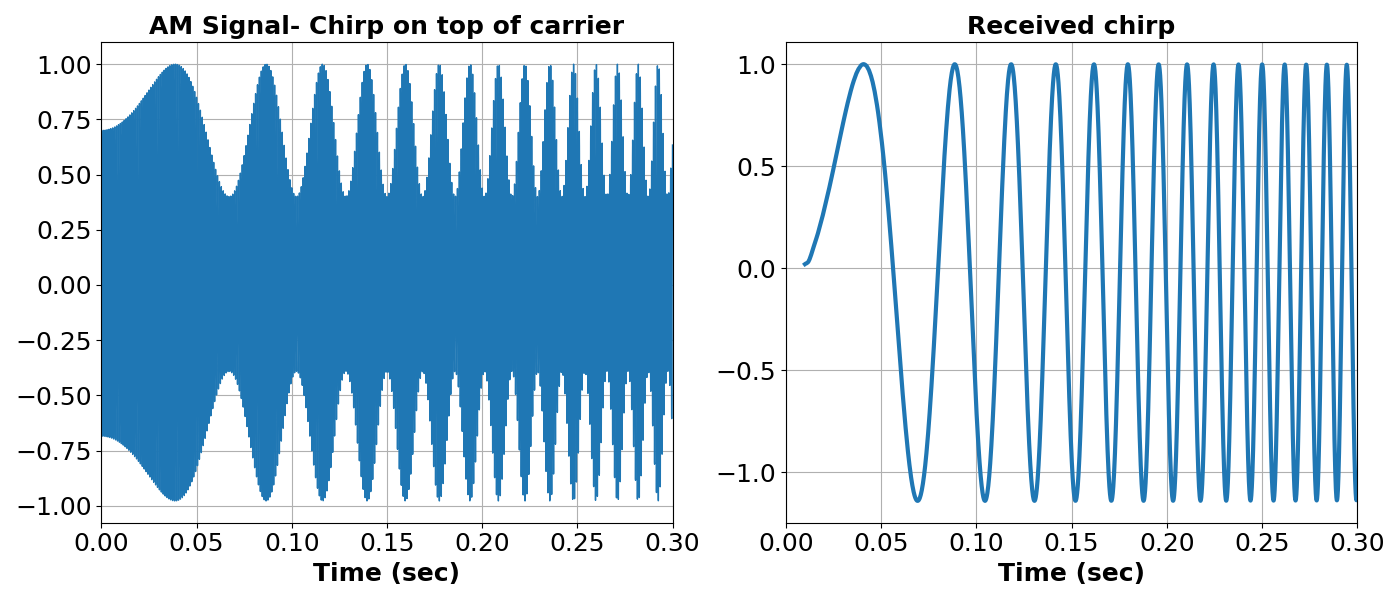}
		\caption{AM Chirp (left) and reconstructed chirp (right).}
		\label{am_chirp}
	\end{center}
\end{figure}

Figure \ref{am_chirp} illustrates a simple case of AM chirp. The left side depicts a 1 kHz carrier modulated in amplitude by a chirp signal with $\alpha = 0.7$, $\beta = 0.3$, $f_1 = 0$ and $f_1 = 50$ Hz. After carrier demodulation, filtering and signal reconstruction in the digital domain, the chirp waveform is successfully recovered as depicted on the right side of Figure \ref{am_chirp}.

\begin{figure}[h]
	\begin{center}
		\includegraphics[width=0.95\columnwidth]{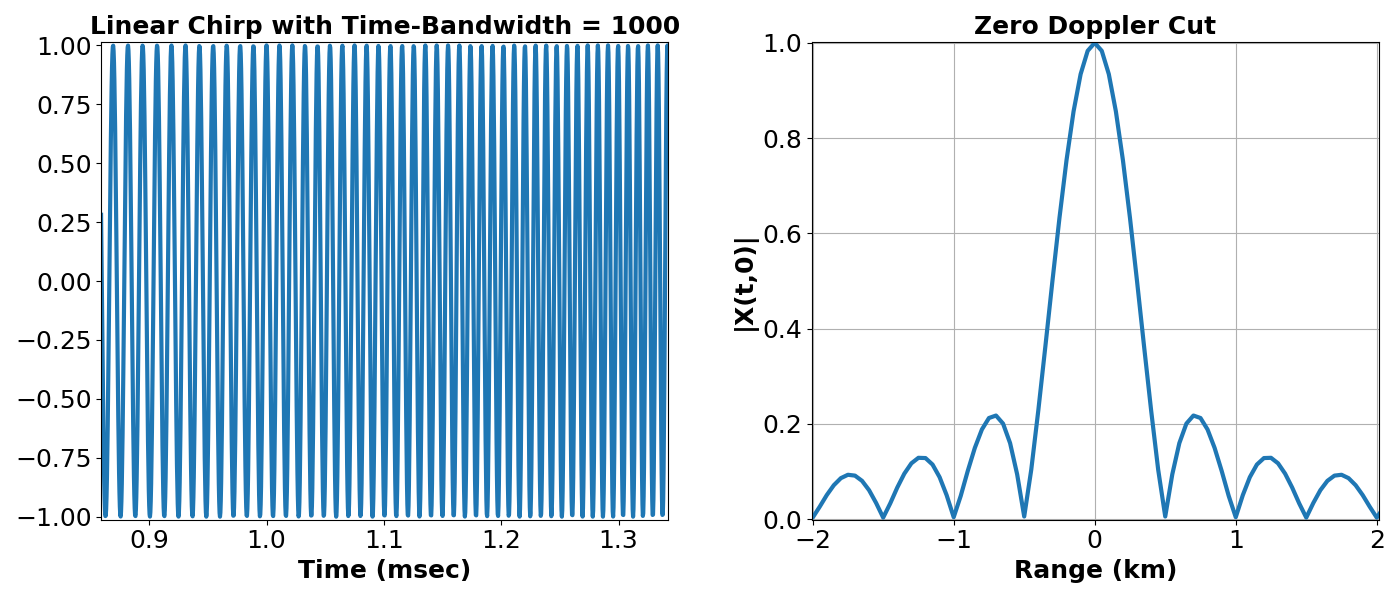}
		\caption{Part of the chirp waveform (left) and zero cut of AF (right).}
		\label{chirp_acf}
	\end{center}
\end{figure}

As said in Section 2, ATLAS has a waveform resolution of 10 ns and a receiver bandwidth of 80~MHz, which enables generation of AM chirps with high compression factors. As an example, Figure \ref{chirp_acf} shows a linear chirp with a compression factor of 1000 inside of the 3.3 ms rectangular pulse. For visualization purposes, only a small fraction of the waveform inside of the pulse is shown. 

A compression factor of 1000 was used to maintain a reasonable simulation time while showing the benefits of using this waveform for pulse compression. As can be seen on the right side of Figure~\ref{chirp_acf}, the range resolution now is approximately 500 m which corresponds to an error of 0.05\% at 1000 km of altitude.

This waveform requires a bandwidth of 300 kHz which is only a small fraction of the capabilities of the system to generate and receive it. This indicates it is possible to synthesize waveforms with even higher compression factors.

\subsection{Advanced Waveform Design}
In the particular scenario where certain selected man-made targets are of interest, the transmitted waveform can be further optimized \cite{PMIRS2017}. The basic idea is to change the signal used as reference for detection in such a way that it is adapted to the predominant scatterers of the targets, resulting in a better matched filtering which ultimately leads to better SNR and pulse compression.

ATLAS also permits the testing of waveforms typically used on Noise Radar Technology \cite{PMAESM2020}, being able to transmit a virtually infinite set of waveforms produced via a set of random realizations, with the real-time configuration of the correspondent matched filter.
This enables the use of the radar in critical environments where interception and jamming robustness is a requisite. Additionally, mutual interference between two radars that occupy the same transmit spectral band can be made negligible~\cite{PMAESM2020} \cite{PMSENSORS2020}.

It is also possible to implement pulse-to-pulse diversity in ATLAS by loading a different waveform during the idle time of the previous pulse. Trains of diverse pulses can lead to reduction of the range sidelobes of the autocorrelation function. The Golay complementary sequences, for example, can be used to phase-modulate the carrier wave leading to trains of complementary pulses. The sum of the autocorrelation of complementary pulses is zero except for the zero shift, leading to an impulse like response \cite{levanon_golay}.

Since the system can be used in passive bistatic mode, it will also be used to test techniques for improved detection by illumination matching on receive such as the one published in \cite{PMEUSAR2018}.

\section{Conclusions and Future Work}
In this work we started by doing a review on the current tracking radar systems deployed around the world as well as the networks responsible for SSA and SST tasks (Section 1). Next, we explained the necessity of developing a tracking radar in Portugal and presented the current one in development, ATLAS. An extensive technical description of the system was provided (Section 2) and simulation studies on its expected performance were developed and discussed (Section 3). Since waveform design plays a big role in radar systems, we gave a description of some interesting waveforms that can be used by the system in order to increase the tracking quality (Section 4).

The system is expected to track targets up to 1000 km in distance with a minimum RCS of 10 cm\textsuperscript{2}, with an average of 500-700 objects passing hourly in the visible sky. With proper waveform design and processing, it is possible to achieve measurements at 1000 km of range with less than 0.05\% error.

The ATLAS pilot radar was developed with cutting edge hardware technology and uses a highly modular architecture with processing fully in the digital domain. This enables a cost reduction in the development and maintenance of the system and provides a platform for research and innovation in the radar field. We also indicated several emerging fields that can be tested and developed using it, such as matched illumination, noise radar and cognitive radar (Section 4.2).

Since the system is at an early stage, several tests and upgrades are envisioned:
\begin{itemize}
    \item Calibration campaigns for the antenna motorized tracking system;
    \item Operational testing in real scenarios;
    \item Incorporation of the system into an operational SST network;
    \item Addition of a waveform generator with In-Phase and Quadrature Modulation;
\end{itemize}
\vspace{6pt} 



\authorcontributions{
conceptualization, João Pandeirada, Miguel Bergano, Domingos Barbosa and Paulo Marques; investigation and writing, João Pandeirada, Miguel Bergano, João Neves and Paulo Marques; simulations, results and visualization, João Pandeirada, Miguel Bergano, Bruno Coelho and Valério Ribeiro; project administration, Miguel Bergano and Domingos Barbosa; funding acquisition, Domingos Barbosa and Valério Ribeiro.}

\funding{The team acknowledges financial support from the Aga Khan Development Network and the Fundação
para a Ciência e a Technologia, Portugal, for the Science and Technology Cooperation DOPPLER - DevelOpment
of PaloP knowLEdge in Radioastronomy, project number 333197717 . V.A.R.M.R. acknowledges financial
support from the Fundacão para a Ciência e Tecnologia (FCT) in the form of an exploratory project of reference
IF/00498/2015 and PHOBOS, POCI-01-0145- FEDER-029932, funded by Programa Operacional Competitividade e Internacionalizacão (COMPETE 2020) and FCT, Portugal. The team acknowledges
financial support from Enabling Green E-science for the Square Kilometre Array Research Infrastructure
(ENGAGE-SKA), POCI-01-0145-FEDER-022217, funded by COMPETE 2020 and FCT, Portugal and from the European Commission H2020 Programme under the grant agreement 2-3SST2018-20.}

\acknowledgments{The team acknowledges the European Commission Horizon 2020 Programme under the Space Surveillance and Tracking (SST) grant agreement 2-3SST2018-20.}

\conflictsofinterest{The authors declare no conflict of interest.} 

\reftitle{References}




\begin{thebibliography}{-------}
	\providecommand{\natexlab}[1]{#1}
	
	\bibitem[{Statista}()]{statista}
	{Statista}.
	\newblock Number of satellites in orbit by major country as of March 31, 2020.
	\newblock
	\url{https://www.statista.com/statistics/264472/number-of-satellites-in-orbit-by-operating-country/#statisticContainer}.
	\newblock [Accessed January, 8, 2021].
	
	\bibitem[{United Nations Office for Outer Space Affairs}()]{outerspace}
	{United Nations Office for Outer Space Affairs}.
	\newblock Treaty on Principles Governing the Activities of States in the
	Exploration and Use of Outer Space, including the Moon and Other Celestial
	Bodies.
	\newblock
	\url{https://www.unoosa.org/oosa/en/ourwork/spacelaw/treaties/introouterspacetreaty.html}.
	\newblock [Accessed January, 8, 2021].
	
	\bibitem[Skolnik(2001)]{skolnik2001}
	Skolnik, M.L.
	\newblock {\em Introduction to RADAR Systems}; Tata McGraw-Hill,  2001.
	
	\bibitem[Chen(2004)]{Chen2004}
	Chen, W.K.
	\newblock {\em The Electrical Engineering Handbook}; Elsevier Academic Press,
	2004.
	
	\bibitem[Galati(2016)]{Galati2016}
	Galati, G.
	\newblock {\em 100 Years of Radar}; Switzerland: Springer,  2016.
	
	\bibitem[Ingwersen and Lemnios(2000)]{Lemnios2000}
	Ingwersen, P.; Lemnios, W.
	\newblock Radars for Ballistic Missile Defense Research,.
	\newblock {\em LINCOLN LABORATORY JOURNAL} {\bf 2000}, {\em 2},~22.
	
	\bibitem[Schrogl \em{et~al.}(2015)Schrogl, Hays, Robinson, Moura, and
	Gianopapa]{Gianopapa2015}
	Schrogl, K.U.; Hays, P.; Robinson, J.; Moura, D.; Gianopapa, C.
	\newblock {\em Handbook of Space Security}; New York: Springer,  2015.
	
	\bibitem[{Thales Group}()]{Thales2020}
	{Thales Group}.
	\newblock Stir - Tracking and illumination radar.
	\newblock
	\url{https://www.thalesgroup.com/en/stir-tracking-and-illumination-radar}.
	\newblock [Accessed September, 17, 2020].
	
	\bibitem[SKA()]{SKA2020}
	SKA.
	\newblock SQUARE KILOMETRE ARRAY.
	\newblock \url{https://www.skatelescope.org/}.
	\newblock [Accessed September, 18, 2020].
	
	\bibitem[eoPortal()]{EOPORTAL2020}
	eoPortal.
	\newblock Space Fence for SSA (Space Situational Awareness) Services.
	\newblock
	\url{https://directory.eoportal.org/web/eoportal/satellite-missions/s/space-fence}.
	\newblock [Accessed September, 18, 2020].
	
	\bibitem[Stokely \em{et~al.}(2006)Stokely, Foster, Stansbery, Benbrook, and
	Juarez]{Stokely2006}
	Stokely, C.L.; Foster, J.L.; Stansbery, E.G.; Benbrook, J.R.; Juarez, Q.
	\newblock Haystack and HAX Radar Measurements of the Orbital Debris Environment
	{\bf 2006}.
	\newblock p. 242.
	
	\bibitem[Murray \em{et~al.}(2019)Murray, Miller, Matney, and
	Kennedy]{Kennedy2019}
	Murray, J.; Miller, R.; Matney, M.; Kennedy, T.
	\newblock Orbital Debris Radar Measurements from the Haystack Ultra-wideband
	Satellite Imaging Radar (HUSIR): 2014-2017.
	\newblock {\em First International Orbital Debris Conference} {\bf 2019}.
	
	\bibitem[Du(2017)]{china}
	Du, R.
	\newblock China's approach to space sustainability: Legal and policy analysis.
	\newblock {\em Space Policy} {\bf 2017}, {\em 42},~8--16.
	
	\bibitem[Prasad(2005)]{india}
	Prasad, M.Y.S.
	\newblock Technical and legal issues surrounding space debris— India’s
	position in the UN.
	\newblock {\em Space Policy} {\bf 2005}, {\em 21},~243--249.
	
	\bibitem[Adimurthy and Ganeshan(2006)]{india1}
	Adimurthy, V.; Ganeshan, A.
	\newblock Space debris mitigation measures in India.
	\newblock {\em Acta Astronautica} {\bf 2006}, {\em 58},~168 -- 174.
	\newblock
	doi:{\changeurlcolor{black}\href{https://doi.org/https://doi.org/10.1016/j.actaastro.2005.09.002}{\detokenize{https://doi.org/10.1016/j.actaastro.2005.09.002}}}.
	
	\bibitem[Froehlich \em{et~al.}(2020)Froehlich, Alonso, Soria, and
	Marchi]{latinamerica}
	Froehlich, A.; Alonso, D.; Soria, A.; Marchi, E.D.
	\newblock {\em Space Supporting Latin America, Latin America's Emerging Space
		Middle Power}; Springer,  2020; pp. 197--273.
	\newblock
	doi:{\changeurlcolor{black}\href{https://doi.org/10.1007/978-3-030-38520-0}{\detokenize{10.1007/978-3-030-38520-0}}}.
	
	\bibitem[Nogueira \em{et~al.}(2012)Nogueira, Andrei, Neto, Tang, Li, Mao,
	Penna, Teixeira, Yu, Neto, and Messias]{chinabrazil}
	Nogueira.; Andrei, E.; Neto, A.; Tang, D.; Li, Z.; Mao, Y.; Penna, Y.;
	Teixeira, J.; Yu, R.; Neto, Y.; Messias.
	\newblock The China-Brazil Program of Space Debris Monitoring.
	\newblock {\em 39th COSPAR Scientific Assembly} {\bf 2012}, {\em
		E1.6-2-12},~1381.
	
	\bibitem[{DEFENSE INTELLIGENCE AGENCY (DIA)}(2019)]{DIA2019}
	{DEFENSE INTELLIGENCE AGENCY (DIA)}.
	\newblock {\em Challenges to Security in Space};  2019; p.~46.
	
	\bibitem[{Mehrholz}(1997)]{Mehrholz1997}
	{Mehrholz}, D.
	\newblock {Radar observations in low earth orbit}.
	\newblock {\em Advances in Space Research} {\bf 1997}, {\em 19},~203--212.
	\newblock
	doi:{\changeurlcolor{black}\href{https://doi.org/10.1016/S0273-1177(97)00002-1}{\detokenize{10.1016/S0273-1177(97)00002-1}}}.
	
	\bibitem[Mehrholz \em{et~al.}(1999)Mehrholz, Leushacke, and Jehn]{Jehn1999}
	Mehrholz, D.; Leushacke, L.; Jehn, R.
	\newblock {COBEAM-1796 Experiment}.
	\newblock {\em Advances in Space Research} {\bf 1999}, {\em 23},~23--32.
	
	\bibitem[Muntoni \em{et~al.}(2017)Muntoni, Schirru, Pisanu, Montisci, Valente,
	Gaudiomonte, Serra, Urru, Ortu, and Fanti]{Muntoni2017}
	Muntoni, G.; Schirru, L.; Pisanu, T.; Montisci, G.; Valente, G.; Gaudiomonte,
	F.; Serra, G.; Urru, E.; Ortu, P.; Fanti, A.
	\newblock Space Debris Detection in Low Earth Orbit with the Sardinia Radio
	Telescope.
	\newblock {\em Electronics} {\bf 2017}, {\em 6}.
	\newblock
	doi:{\changeurlcolor{black}\href{https://doi.org/10.3390/electronics6030059}{\detokenize{10.3390/electronics6030059}}}.
	
	\bibitem[{European Union}(2014)]{EUNION2014}
	{European Union}.
	\newblock DECISION No 541/2014/EU OF THE EUROPEAN PARLIAMENT AND OF THE COUNCIL
	of 16 April 2014 establishing a Framework for Space Surveillance and Tracking
	Support,  2014.
	
	\bibitem[{EU SST}()]{eusst}
	{EU SST}.
	\newblock EUSST-EUROPEAN SPACE SURVEILLANCE AND TRACKING PROJECTS.
	\newblock \url{https://www.eusst.eu}.
	
	\bibitem[Eshbaugh \em{et~al.}(2014)Eshbaugh, Morrison, Hoen, Hiett, and
	Benitz]{husir}
	Eshbaugh, J.V.; Morrison, R.L.; Hoen, E.W.; Hiett, T.C.; Benitz, G.R.
	\newblock {HUSIR Signal Processing}.
	\newblock {\em Lincoln Laboratory Journal} {\bf 2014}, {\em 21},~115--134.
	
	\bibitem[{Rejto}(2000)]{rosa}
	{Rejto}, S.
	\newblock Radar open systems architecture and applications.
	\newblock  Record of the IEEE 2000 International Radar Conference [Cat. No.
	00CH37037],  2000, pp. 654--659.
	\newblock
	doi:{\changeurlcolor{black}\href{https://doi.org/10.1109/RADAR.2000.851911}{\detokenize{10.1109/RADAR.2000.851911}}}.
	
	\bibitem[Fonseca \em{et~al.}(2006)Fonseca, Barbosa, Cupido, Mourao, Santos,
	Smoot, and Tello]{rui}
	Fonseca, R.; Barbosa, D.; Cupido, L.; Mourao, A.; Santos, D.; Smoot, G.; Tello,
	C.
	\newblock Site evaluation and RFI spectrum measurements in Portugal at
	the481frequency range 0.408–10 GHz for a GEM polarized galactic radio
	emission experiment,.
	\newblock {\em New Astronomy - NEW ASTRON} {\bf 2006}, {\em 11},~551--556.
	
	\bibitem[{Cree (2018)}()]{cree}
	{Cree (2018)}.
	\newblock CGHV59350 Data Sheet, 350 W, 5200-5900 MHz GaN HEMT for C-Band Radar.
	\newblock
	\url{https://www.wolfspeed.com/downloads/dl/file/id/463/product/174/cghv59350.pdf}.
	\newblock [accessed May 29, 2018].
	
	\bibitem[{Simons} \em{et~al.}(2019){Simons}, {Wintucky}, and
	{Waldstein}]{simons}
	{Simons}, R.N.; {Wintucky}, E.G.; {Waldstein}, S.W.
	\newblock A Novel Reconfigurable GaN Based Fully Solid-State Microwave Power
	Module for Communications/Radar Applications.
	\newblock  2019 IEEE Aerospace Conference,  2019, pp. 1--7.
	\newblock
	doi:{\changeurlcolor{black}\href{https://doi.org/10.1109/AERO.2019.8741973}{\detokenize{10.1109/AERO.2019.8741973}}}.
	
	\bibitem[{General Microwave}()]{pinmod}
	{General Microwave}.
	\newblock Series D195 Octave-Band PIN Diode Attenuator/Modulators.
	
	\bibitem[{Wolfspeed Cree}()]{gan1}
	{Wolfspeed Cree}.
	\newblock CGHV59070 70 W, 4.4 - 5.9 GHz, 50 V, RF Power GaN HEMT.
	\newblock [Rev 2.2 - April 2020].
	
	\bibitem[{MTI-Milliren Technologies, Inc.}()]{ocxo}
	{MTI-Milliren Technologies, Inc.}.
	\newblock 260 Series OCXO.
	\newblock [Mar 2015].
	
	\bibitem[{Advantech}()]{sbc}
	{Advantech}.
	\newblock MIO-2360 IntelR©Pentium N4200/Celeron N3350/AtomTME3900 series
	Pico-ITX SBC.
	\newblock [Last updated: 14-Sep-2018].
	
	\bibitem[Skolnik(2008)]{skolnik}
	Skolnik, M.L.
	\newblock {\em {{"An Introduction and Overview of Radar", in "Radar
				Handbook"}}}; McGraw-Hill,  2008.
	
	\bibitem[{International Telecommunication Union}()]{itur}
	{International Telecommunication Union}.
	\newblock {"Propagation data and prediction methods required for the design of
		Earth-space telecommunication systems", in "Recommendation ITU-R"}.
	
	\bibitem[{Riché} \em{et~al.}(2014){Riché}, {Méric}, {Baudais}, and
	{Pottier}]{ofdm_interleaving}
	{Riché}, V.; {Méric}, S.; {Baudais}, J.; {Pottier}, î
	\newblock Investigations on OFDM Signal for Range Ambiguity Suppression in SAR
	Configuration.
	\newblock {\em IEEE Transactions on Geoscience and Remote Sensing} {\bf 2014},
	{\em 52},~4194--4197.
	\newblock
	doi:{\changeurlcolor{black}\href{https://doi.org/10.1109/TGRS.2013.2280190}{\detokenize{10.1109/TGRS.2013.2280190}}}.
	
	\bibitem[{USSPACECOM}()]{spacetrack}
	{USSPACECOM}.
	\newblock Space-Track.
	\newblock \url{https://www.space-track.org/}.
	
	\bibitem[{Satellogic SA}()]{pythonpred}
	{Satellogic SA}.
	\newblock Orbit Predictor.
	\newblock \url{https://github.com/satellogic/orbit-predictor}.
	\newblock [Accessed December, 30, 2020].
	
	\bibitem[N.Levanon and E.Mozeson(2004{\natexlab{a}})]{levanon_amb}
	N.Levanon.; E.Mozeson.
	\newblock {\em {{"Ambiguity Function", in "Radar Signals"}}}; John Wiley \& Sons, Inc.,  2004; pp. 34--52.
	
	\bibitem[N.Levanon and E.Mozeson(2004{\natexlab{b}})]{levanon_matched}
	N.Levanon.; E.Mozeson.
	\newblock {\em {{"Matched Filter", in "Radar Signals"}}}; John Wiley \& Sons, Inc.,  2004; pp. 20--33.
	
	\bibitem[Skolnik(2008)]{skolnikpulsecomp}
	Skolnik, M.L.
	\newblock {\em {{"Pulse Compression Radar", in "Radar Handbook"}}};
	McGraw-Hill,  2008.
	
	\bibitem[N.Levanon and E.Mozeson(2004)]{levanon_barker}
	N.Levanon.; E.Mozeson.
	\newblock {\em {{"Phase-Coded Pulse", in "Radar Signals"}}}; John Wiley \& Sons, Inc.,  2004; pp. 100--167.
	
	\bibitem[Dzvonkovskaya and Rohling(2008)]{dzvonkovskaya2008}
	Dzvonkovskaya, A.; Rohling, H.
	\newblock Long binary phase codes with good autocorrelation properties.
	\newblock  2008 International Radar Symposium,  2008, pp. 1--4.
	\newblock
	doi:{\changeurlcolor{black}\href{https://doi.org/10.1109/IRS.2008.4585750}{\detokenize{10.1109/IRS.2008.4585750}}}.
	
	\bibitem[N.Levanon and E.Mozeson(2004)]{levanon_lfm}
	N.Levanon.; E.Mozeson.
	\newblock {\em {{"Basic Radar Signals", in "Radar Signals"}}}; John Wiley \& Sons, Inc.,  2004; pp. 53--73.
	
	\bibitem[{Marques}(2017)]{PMIRS2017}
	{Marques}, P.A.
	\newblock Noise radar detection optimized for selected targets.
	\newblock  2017 18th International Radar Symposium (IRS),  2017, pp. 1--9.
	\newblock
	doi:{\changeurlcolor{black}\href{https://doi.org/10.23919/IRS.2017.8008226}{\detokenize{10.23919/IRS.2017.8008226}}}.
	
	\bibitem[{K. Savci et al}(2020)]{PMAESM2020}
	{K. Savci et al}.
	\newblock Noise Radar—Overview and Recent Development.
	\newblock {\em IEEE Aerospace and Electronic498Systems Magazine} {\bf 2020},
	{\em 35},~8--20.
	\newblock
	doi:{\changeurlcolor{black}\href{https://doi.org/10.1109/MAES.2020.2990591}{\detokenize{10.1109/MAES.2020.2990591}}}.
	
	\bibitem[Palo \em{et~al.}(2020)Palo, Galati, Pavan, Wasserzier, and
	Savci]{PMSENSORS2020}
	Palo, F.D.; Galati, G.; Pavan, G.; Wasserzier, C.; Savci, K.
	\newblock Introduction to Noise Radar and Its Waveforms.
	\newblock {\em Sensors} {\bf 2020}, {\em 20}.
	\newblock
	doi:{\changeurlcolor{black}\href{https://doi.org/10.3390/s20185187}{\detokenize{10.3390/s20185187}}}.
	
	\bibitem[N.Levanon and E.Mozeson(2004)]{levanon_golay}
	N.Levanon.; E.Mozeson.
	\newblock {\em {{"Coherent train of diverse pulses", in "Radar Signals"}}};
	John Wiley \& Sons, Inc.,  2004; pp. 262--270.
	
	\bibitem[{Marques}(2018)]{PMEUSAR2018}
	{Marques}, P.A.
	\newblock More Illumination for Passive Radar.
	\newblock  European Conf. on Synthetic Aperture Radar,  2018, Vol.~1.
	
\end{thebibliography}



\end{document}